\documentclass[usenatbib, usegraphicx]{mn2e}

\newlength{\colwidth}
\newcommand{\cm}{{\rm cm}} \newcommand{\s}{{\rm s}}
 \newcommand{\K}{{\rm K}}

\title[Cooling of enriched, photo-ionized gas]{The effect of photo-ionization on the cooling rates of
  enriched, astrophysical plasmas} 
\author[Wiersma, Schaye, \& Smith]{Robert P.C. Wiersma,$^1$\thanks{E-mail:
wiersma@strw.leidenuniv.nl} Joop~Schaye$^1$ and Britton D. Smith$^2$
  \\ $^1$Leiden Observatory, Leiden
  University, P.O. Box 9513, 2300 RA Leiden, The Netherlands\\
$^2$Center for Astrophysics \& Space Astronomy, Department of
  Astrophysical \& Planetary Sciences,\\ University 
of Colorado, Boulder, CO, 80309, USA}

\begin{document}

\pagerange{\pageref{firstpage}--\pageref{lastpage}} \pubyear{2008}

\maketitle

\label{firstpage}

\begin{abstract}

Radiative cooling is central to a wide range of astrophysical
problems. Despite its importance, cooling rates are generally computed
using very restrictive assumptions, such as collisional ionization
equilibrium and solar relative abundances. We simultaneously relax
both assumptions and investigate the effects of photo-ionization of
heavy elements by the meta-galactic UV/X-ray background and of
variations in relative abundances on the cooling rates of optically
thin gas in ionization equilibrium. We find that photo-ionization by
the meta-galactic background radiation reduces the net cooling rates by up
to an order of magnitude for gas densities and temperatures typical of
the shock-heated intergalactic medium and proto-galaxies ($10^4\,\K
\la T \la 10^6\,\K$, $\rho / \left<\rho\right> \la 100$). In addition, photo-ionization
changes the relative contributions of different elements to the
cooling rates. We conclude that photo-ionization by the ionizing
background and heavy elements both need to be taken into account in
order for the cooling rates to be correct to order of
magnitude. Moreover, if the rates need to be known to better than a
factor of a few, then departures of the relative abundances from solar
need to be taken into account. We propose a method to compute cooling
rates on an element-by-element basis by interpolating pre-computed
tables that take photo-ionization into account. We provide such tables
for a popular model of the evolving UV/X-ray background radiation,
computed using the photo-ionization package \textsc{cloudy}.

\end{abstract}


\section{Introduction}
\label{sec:intro}
Dissipation of energy via radiative cooling plays a central role in
many different astrophysical contexts. In general the cooling rate
depends on a large number of parameters, such as the gas density,
temperature, chemical composition, ionization balance, and the
radiation field. In the absence of radiation, however, the equilibrium
ionization balance depends only on the temperature. In that case the
cooling rate in the low density regime, which is dominated by collisional 
processes, is simply proportional to the
gas density squared, for a given composition. Thus, the cooling rates for a plasma in
collisional ionization equilibrium (CIE) can be conveniently tabulated
as a function of the temperature and composition (metallicity) of the
gas
\cite[e.g.,][]{Cox1969,Raymond1976,Shull1982,Gaetz1983,Boehringer1989,Sutherland1993,Landi1999,Benjamin2001,Gnat2007,Smith2008},
and such tables are widely used for a large variety of problems.

Although it is convenient to ignore radiation when calculating cooling
rates, radiation is generally important for the thermal and ionization
state of astrophysical plasmas. For example, \cite{Efstathiou1992}
investigated the effect of the extragalactic UV background on the
cooling rates for gas of primordial composition (in practice this
means a pure H/He plasma) and found that including photo-ionization
can suppress the cooling rates of gas in the temperature range $T\sim
10^4 - 10^5\,\K$ by a large factor. Although the effects of radiation
are often taken into account for gas of primordial composition,
photo-ionization of heavy elements is usually ignored in the
calculation of cooling rates (but see \citealt{Leonard1998,Benson2002}). 

In this paper we will investigate the dependence of cooling rates of
gas enriched with metals on the presence of ionizing radiation,
focusing on the temperature range $T\sim 10^4 - 10^8\,\K$ and
optically thin plasmas. We will show that, as is the case for gas of
primordial composition \cite[][]{Efstathiou1992}, photo-ionization can
suppress the metallic cooling rates by a large factor. Moreover, the
suppression of the cooling rate is significant up to much higher
temperatures than for the primordial case.

We will also investigate the relative contributions of various
elements to the cooling rates. If the relative abundances are similar
to solar, then oxygen, neon, and iron dominate the cooling in the
temperature range $T \sim 10^4 - 10^7\,\K$. However, we will show that
the relative contributions of different elements to the cooling rate
are sensitive to the presence of ionizing radiation.

Although we will illustrate the results using densities and radiation
fields that are relevant for studies of galaxy formation and the
intergalactic medium (IGM), the conclusion that photo-ionization
significantly reduces the cooling rates of enriched gas is valid for a
large variety of astrophysical problems. For example, for $T \sim
10^5\,\K$ and $T\sim 10^6\,\K$ the reduction of metal-line cooling rates is
significant as long as the dimensionless 
ionization parameter\footnote{$U \equiv \Phi_{\rm H}/(n_{\rm H}c)$,
  where $\Phi_{\rm H}$ is the flux of hydrogen ionizing photons
  (i.e.,\ photons per unit area and time), $n_{\rm H}$ is the total
  hydrogen number density and $c$ is the speed of
  light.} $U \ga 10^{-3}$ and $U \ga 10^{-1}$, respectively. We will
focus on the temperature 
range $10^4 - 10^8\,\K$ because gas in this temperature range is
usually optically thin and because the effects of photo-ionization are
generally unimportant at higher temperatures.

Tables containing cooling rates and several other useful quantities as
a function of density, temperature, redshift, and composition,
appropriate for gas exposed to the models for the evolving
meta-galactic UV/X-ray background of \cite{Haardt2001} are available
on the following web site:
\texttt{http://www.strw.leidenuniv.nl/WSS08/}. The web site also
contains a number of videos that illustrate the dependence of the
cooling rates on various parameters.

This paper is organized as follows. In Section~\ref{sec:method} we
present our method for calculating element-by-element cooling rates
including photo-ionization and we compare the limiting case of CIE to
results taken from the literature. Section~\ref{sec:photometals} shows
how metals and ionizing radiation affect the cooling rates.
Section~\ref{sec:whim} demonstrates the importance for the
low-redshift shock-heated IGM, which is thought to contain most of the
baryons. In this section we also illustrate the effect of changing the
intensity and spectral shape of the ionizing radiation. We investigate
the effect of photo-ionization on the relative contributions of
individual elements in Section~\ref{sec:relabund} and we summarize and
discuss our conclusions in Section~\ref{sec:discussion}. 

Throughout this paper we use the cosmological parameters from 
\cite{Komatsu2008}: ($\Omega_{\rm m}, \Omega_{\Lambda}, \Omega_{\rm
  b}, h) = (0.279, 0.721, 0.0462, 0.701$) and a primordial helium
mass fraction $X_{\rm He} = 0.248$. Densities will be expressed both
as proper hydrogen number densities $n_{\rm H}$ and density
contrasts $\delta \equiv \rho_{\rm b}/\left <\rho_{\rm b}\right >-1$,
where $\left < \rho_{\rm b}\right >$ is the cosmic mean baryon
density. The two are related by 
\begin{equation}
n_{\rm H} \approx 1.9 \times 10^{-7}\,\cm^{-3} ~\left (1+\delta\right
)\left (1+z\right )^3 
\left ({X_{\rm H} \over 0.752}\right ).
\end{equation}

\section{Method}
\label{sec:method}
All radiative cooling and heating rates were computed by running large
grids of photo-ionization models using the publicly available
photo-ionization package
\textsc{cloudy}\footnote{\texttt{http://www.nublado.org/}} (version
07.02 of the code last described by \citealt{Ferland1998}).
\textsc{cloudy} contains most of the atomic processes that are thought
to be important in the temperature range of interest here ($T \sim
10^4 - 10^8\,\K$) and the reader is referred to the online
documentation for details about the atomic physics and data used.

The gas was exposed to the cosmic microwave background radiation (CMB)
and the \citet[][hereafter HM01]{Haardt2001}
model\footnote{\texttt{http://pitto.mib.infn.it/$\sim$haardt/refmodel.html}}
for the UV/X-ray background radiation from galaxies (assuming a 10
percent escape fraction for H-ionizing photons) and quasars. We
assumed the gas to be dust-free, optically thin and in ionization
equilibrium. We discuss the limitations and the effects of the last
two assumptions in section~\ref{sec:discussion}. All cooling rates are
tabulated as a 
function of $\log n_{\rm H}$ (total hydrogen number density), $\log T$
(temperature), and $z$ (redshift).

\begin{table}
\caption{Default \textsc{cloudy} solar abundances}
\label{tab-abund}
\centering
\begin{tabular}{ccc}
\hline\hline Element & $n_i/n_{\rm H}$ & Mass Fraction\\ 
\hline 
H & 1 & 0.7065 \\ 
He & 0.1 & 0.2806 \\ 
C & $2.46\times 10^{-4}$ & $2.07\times 10^{-3}$ \\ 
N & $8.51\times 10^{-5}$ & $8.36\times 10^{-4}$ \\ 
O & $4.90\times 10^{-4}$ & $5.49\times 10^{-3}$ \\ 
Ne & $1.00\times 10^{-4}$ & $1.41\times 10^{-3}$ \\ 
Mg & $3.47\times 10^{-5}$ & $5.91\times 10^{-4}$ \\ 
Si & $3.47\times 10^{-5}$ & $6.83\times 10^{-4}$ \\ 
S & $1.86\times 10^{-5}$ & $4.09\times 10^{-4}$ \\ 
Ca & $2.29\times 10^{-6}$ & $6.44\times 10^{-5}$ \\ 
Fe & $2.82\times 10^{-5}$ & $1.10\times 10^{-3}$ \\ 
\hline
\end{tabular}
\end{table} 

When studying the effect of changes in the relative abundances of
elements, we compute the cooling rates on an element-by-element
basis. The cooling rate $\Lambda_i$ (in ${\rm
  erg} \,{\rm s}^{-1}\, {\rm cm}^{-3}$) due to element $i$, where element $i$ is
heavier than helium, is defined as the difference between the cooling
rate computed using all elements (assuming solar abundances) and the
cooling rate computed after setting the abundance of element $i$ to
zero, while keeping all other abundances (i.e., number densities
relative to H) the same. This is a valid approximation provided
element $i$ does not contribute significantly to the free electron
density, which is the case for all elements heavier than helium and if
the metallicity $Z \la Z_\odot$.

The combined contributions from hydrogen and helium are computed by
interpolating in the four dimensions $\log n_{\rm H}$, $\log T$, $z$,
and $n_{\rm He}/n_{\rm H}$ from tables of \textsc{cloudy} models that
contain only H and He.

Thus, the total net cooling rate,
\begin{equation}
\Lambda = \Lambda_{\rm H,He} + \sum_{i>{\rm He}} \Lambda_i,
\end{equation}
could be obtained from
\begin{equation} 
\Lambda = \Lambda_{\rm H,He} + \sum_{i>{\rm He}} {n_i/n_{\rm H} \over (n_i/n_{\rm
    H})_\odot}\Lambda_{i,\odot},
\label{eq:badcoolmethod}
\end{equation}
where $(n_i/n_{\rm H})_\odot$ is the solar
abundance of element $i$ (the default \textsc{cloudy} solar abundances
are given in Table~\ref{tab-abund}) and $\Lambda_{i,\odot}$ is the
contribution of heavy element $i$ to the radiative cooling rate for
solar abundances, which we have tabulated as a function of $\log
n_{\rm H}$, $\log T$, and $z$. Note that we use $\Lambda$ to denote the cooling rate per unit volume (${\rm
  erg} \,{\rm s}^{-1}\, {\rm cm}^{-3}$).

We can, however, do better than equation (\ref{eq:badcoolmethod}) by taking the
dependence of the free electron density on He/H, into account 
(in equation (\ref{eq:badcoolmethod}), the electron density is implicitly assumed to be 
that corresponding to solar abundances - $n_{{\rm He}}/n_{{\rm H}} = 0.1$). Since cooling
rates due to metals are dominated by collisions between ions and free
electrons, $\Lambda_i$ scales as the product of the free electron and
ion densities, $\Lambda_i \propto n_{\rm e} n_i$. Hence,
\begin{eqnarray}
\Lambda &=& \Lambda_{\rm H,He} + \sum_{i>{\rm He}} \Lambda_{i,\odot}
\left ({n_{\rm e} \over n_{{\rm e},\odot}}\right ) \left ({n_i \over
  n_{i,\odot}}\right ), \nonumber \\ 
&=& \Lambda_{\rm H,He} + \sum_{i>{\rm He}}
\Lambda_{i,\odot} {n_{\rm e}/n_{\rm H} \over (n_{\rm e}/n_{\rm
    H})_\odot} 10^{\left [i/{\rm H}\right ]}
\label{eq:coolmethod}
\end{eqnarray}
where $10^{[i/{\rm H}]} \equiv (n_i/n_{\rm H}) / (n_i/n_{\rm H})_\odot$ and we used the fact
that $n_{\rm H} = n_{{\rm H},\odot}$ (since we 
tabulate as a function of $n_{\rm H}$). While $(n_{\rm e}/n_{\rm
  H})_\odot$ is obtained by interpolating the solar abundance table
for $n_{\rm e}/n_{\rm H}$ in $(n_{\rm H},T,z)$, $(n_{\rm e}/n_{\rm
  H})$ must be obtained by interpolating in $((n_{\rm He}/n_{\rm
  H}),n_{\rm H},T,z)$. Note that we tabulate the electron density in the absence 
of metals. This is a valid approximation given that heavy elements
only contribute significantly to the free electron density for $Z \gg Z_{\odot}$.

In practice we tabulate $\log \Lambda_{i,\odot}/n_{\rm H}^2$ over the
range $\log[ n_{\rm H}\,(\cm^{-3})] = -8.0$, -7.9, $\ldots$, 0.0 and
$\log[T~(\K)] = 2.00$, 2.02, $\ldots$, 9.00. In addition, the
quantities $\log \Lambda_{{\rm H},{\rm He}}/n_{\rm H}^2$, $n_{\rm
  e}/n_{\rm H}$ and the mean particle mass are all tabulated as a
function of density and temperature for each of the values $n_{\rm
  He}/n_{\rm H} = 0.0787$, 0.0830, 0.0876, 0.0922, 0.970, 0.102, 0.107
which correspond to mass fractions $X_{\rm He}/(X_{\rm H} + X_{\rm
  He}) = 0.238$, 0.248, 0.258, 0.268, 0.278, 0.288, 0.298. Finally,
for each value of the ratio $n_{\rm He}/n_{\rm H}$ we tabulate the
temperature as a function of density and internal energy per unit mass
to enable simulation codes that parametrize thermal energy in terms of
the latter quantity to use the cooling tables. We have computed these
tables for each of the 49 redshifts spanning $z=0-9$ for which the
HM01 models are defined. All tables are in HDF5 format and together
they result in upwards of 232~MB of storage. Reducing the resolution
in $n_{\rm H}$ and $T$ by a factor of two does reduce the accuracy of
interpolated rates significantly\footnote{Reducing the temperature
  resolution by a factor of two roughly doubles the interpolation
  errors near the thermal equilibrium solution, but in this regime the
  cooling times are in any case effectively infinite.}, but reduces
the storage requirements to 61~MB.

We find that for metallicities $Z \la Z_\odot$
equation~(\ref{eq:coolmethod}) closely matches the true cooling rate
(i.e., including all elements) with very good accuracy\footnote{While 
  the error in the total cooling rates is small,
  the relative errors in the contribution of individual elements can
  be larger if the element contributes negligibly to the total cooling
  rate (see e.g.,\ the noise in the Fe contour at $T<10^5\,\K$ in the
  left-hand panel of figure~\ref{fig:elements}). This is because the
  tables for individual elements were computed by taking the
  difference between the total cooling rate for solar abundances
  including and excluding the element. Hence, if the element does not
  contribute significantly to the cooling rate for a particular
  density, temperature, and redshift, then its contribution will be
  computed as the difference between two nearly equal numbers. Because
  this can only occur if the contribution is negligible, we do not
  consider this to be a problem.}  if the following 11 elements are
included: H, He, C, N, O, Ne, Mg, Si, S, Ca, and Fe.

\begin{figure}
\includegraphics[width=84mm]{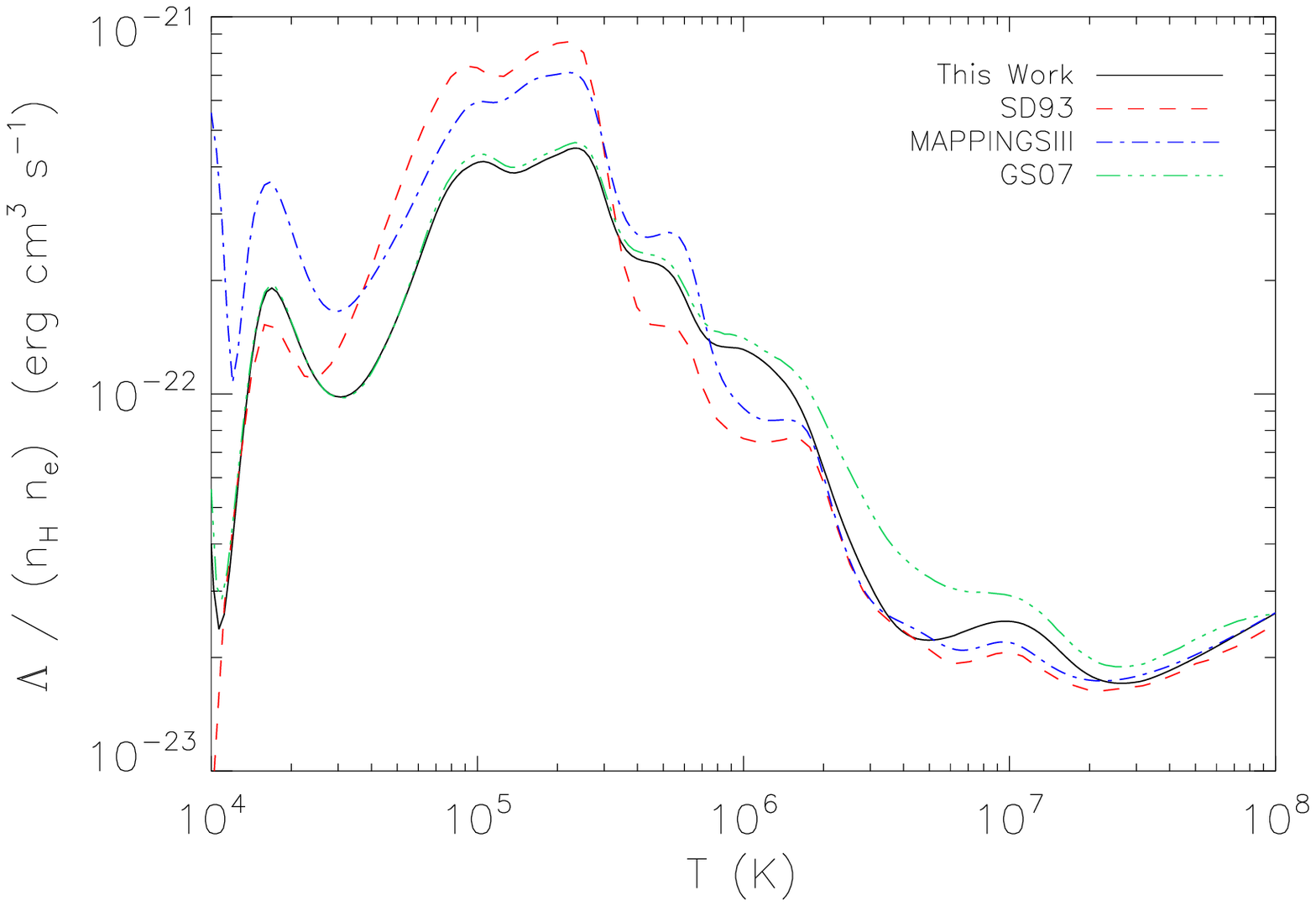}
\caption{Comparison of normalized CIE cooling rates for various
  studies. All curves use the solar abundances of
  \protect\cite{Gnat2007}, which differ somewhat from our default
  solar abundances. Shown are \textsc{cloudy} version 07.02 (solid),
  \protect\cite{Sutherland1993} (dashed), \textsc{mappings iii}
  (dot-dashed), and \protect\cite{Gnat2007}
  (dot-dot-dot-dashed). \textsc{mappings} gives significantly higher
  cooling rates for $T\sim 10^5\,\K$, but the differences are
  typically smaller than a factor of two. Note that for this
  comparison the cooling rates were divided by $n_{\rm H}n_{\rm e}$,
  but that we divide by $n_{\rm H}^2$ in our tables and in
  figure~\protect\ref{fig:elements}. The upturn in the normalized
  cooling rates below $10^4~\K$ is caused by the sharp decrease in
  $n_{\rm e}$ with decreasing temperature.
\label{fig:codecomparison}}
\end{figure}

For redshift $z=0$ and over the full range of densities and
temperatures, the median relative 
errors in the absolute net cooling rates are 0.33\%,
1.6\%, and 6.1\% for $Z = 0.1 Z_{\odot}$, $Z_{\odot}$, and $10
Z_{\odot}$, respectively (we have scaled $n_{{\rm He}}/n_{{\rm H}}$ with 
metallicity). For higher redshifts 
the median errors are smaller than for $z=0$ because Compton cooling
off the CMB, which is modelled accurately, becomes increasingly
important. Hence, even for 
metallicities as extreme as 10 times solar, using
equation~(\ref{eq:coolmethod}) and including 11 elements gives errors of only a
few percent. As we shall see below, this is much smaller than the
differences between different photo-ionization codes. Using 
equation~(\ref{eq:badcoolmethod}) rather than (\ref{eq:coolmethod})
gives similar errors for low metallicities, but the median errors are
a factor 2--3 higher for $Z > 10^{0.5}~Z_\odot$. 

Excluding temperatures within 0.1~dex of the thermal equilibrium
solution (where the relative errors in the net cooling rate, which is
computed as the absolute difference between heating and cooling,
become large because we are subtracting two nearly identical numbers),
the maximum errors in the net cooling rates are 32\%, 29\%, and 39\%
for $Z = 0.1 Z_{\odot}$, $Z_{\odot}$, and $10 Z_{\odot}$,
respectively.  Thus, even for extreme metallicities the maximum
difference between the estimated and true rates is well within a
factor of two.  These maxima are reached near the 0.1~dex exclusion
zone, so the maximum errors for temperatures that differ substantially
from the equilibrium values are much smaller.

We have also tabulated the combined cooling rates of all elements
heavier than helium, assuming solar relative abundances. While scaling
all heavy elements simultaneously reduces the accuracy (relative to
scaling the contribution of each element individually), these tables may be
convenient in the absence of a complete set of elemental abundances. In this
case, the total cooling rate becomes:

\begin{equation}
\Lambda = \Lambda_{\rm H,He} + \Lambda_{Z,\odot} {n_{\rm e}/n_{\rm H} \over (n_{\rm e}/n_{\rm H})_\odot} {Z \over Z_{\odot}}.
\label{eq:Zmethod}
\end{equation}

Although cooling rates including photo-ionization have not yet been
tabulated, CIE cooling rates have been published by various
authors. In Figure~\ref{fig:codecomparison} we compare our CIE results
(i.e., \textsc{cloudy} version 07.02) with those of \cite{Gnat2007}
(who used \textsc{cloudy} version 06.02), \cite{Sutherland1993} (who
used \textsc{mappings}) and also with \textsc{mappings iii}
\citep[version r;][]{Groves2008}.  The calculations generally differ in terms of
the code, the atomic rates, and the solar abundances that were
used. In order to focus the comparison on codes and rates, we used the
solar abundances assumed by \cite{Gnat2007} when computing the cooling
rates for all curves in this figure\footnote{\cite{Sutherland1993}
  only give the cooling rates as a function of the abundance of iron,
  which we interpolated to the corresponding \cite{Gnat2007}
  value.}. Note that these abundances differ somewhat from the solar
abundances that we use in the rest of this paper.

The differences in the cooling rates shown in
Figure~\ref{fig:codecomparison} are typically smaller than 0.3~dex
suggesting that the cooling rates are known to better than a factor of
two, at least for CIE.  The differences are largest for $T\sim
10^5\,\K$, with \textsc{mappings} giving higher cooling rates than
\textsc{cloudy}.

\section{Photo-Ionization, metals, and cooling rates}
\label{sec:photometals}

\begin{figure}
\includegraphics[width=84mm]{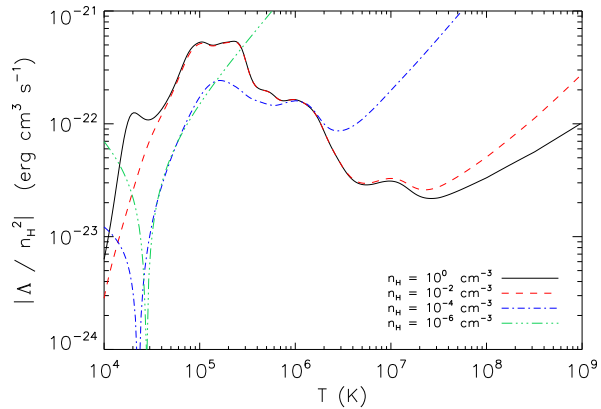}
\caption{Normalized, absolute, net cooling rates ($\vert\Lambda/n_{\rm H}^2\vert$) as a function of
  temperature for solar abundances. Different curves correspond to
  different gas densities, as indicated in the legend. The gas in all
  models is exposed to the $z=3$ CMB and the $z=3$ HM01 UV/X-ray
  background. Note that the cooling rates were 
  divided by $n_{\rm H}^2$, but that we divided by $n_{\rm H}n_{\rm e}$
  in Fig.~\protect\ref{fig:codecomparison}.
\label{fig:photodensity}}
\end{figure}

The cooling rates in photo-ionization equilibrium 
are not proportional to the density squared, as is the case for
CIE. As a consequence, the cooling  
curve changes dramatically with density. This is illustrated in
Fig.~\ref{fig:photodensity} which shows normalized, net cooling, 
rates ($\vert \Lambda/n_{\rm H}^2\vert$),
as a function of temperature for solar abundances. Different
curves correspond to different densities, as indicated
in the legend. 

The calculation includes the CMB and the HM01 background
at redshift three. The $z=3$ HM01 background corresponds to a hydrogen
photo-ionization rate $\Gamma_{12} \equiv \Gamma / 10^{-12}~\s^{-1} =
1.1$ and an ionization parameter $U = 1.6 \times 10^{-5}/n_{\rm
  H}$. We stress that in the vicinity of ionizing sources the 
radiation field may be much more intense, which would enhance its
effect on the cooling rates compared to the results presented here.

As the density decreases
the gas becomes more highly ionized and the cooling peaks due to
collisional excitation of various ions disappear. For very low
densities and high temperatures the normalized cooling rates actually
increase as they become dominated by Compton cooling off the CMB, which
scales as $\Lambda \propto n_{\rm e}T$. For the two lowest densities
shown, the equilibrium temperature, for which the net cooling rate
$\Lambda \rightarrow 0$, exceeds $10^4\,\K$. In these cases, curves 
corresponding to net heating appear to the left of the thermal 
equilibrium temperature. 

\begin{figure*}
\includegraphics[width=84mm]{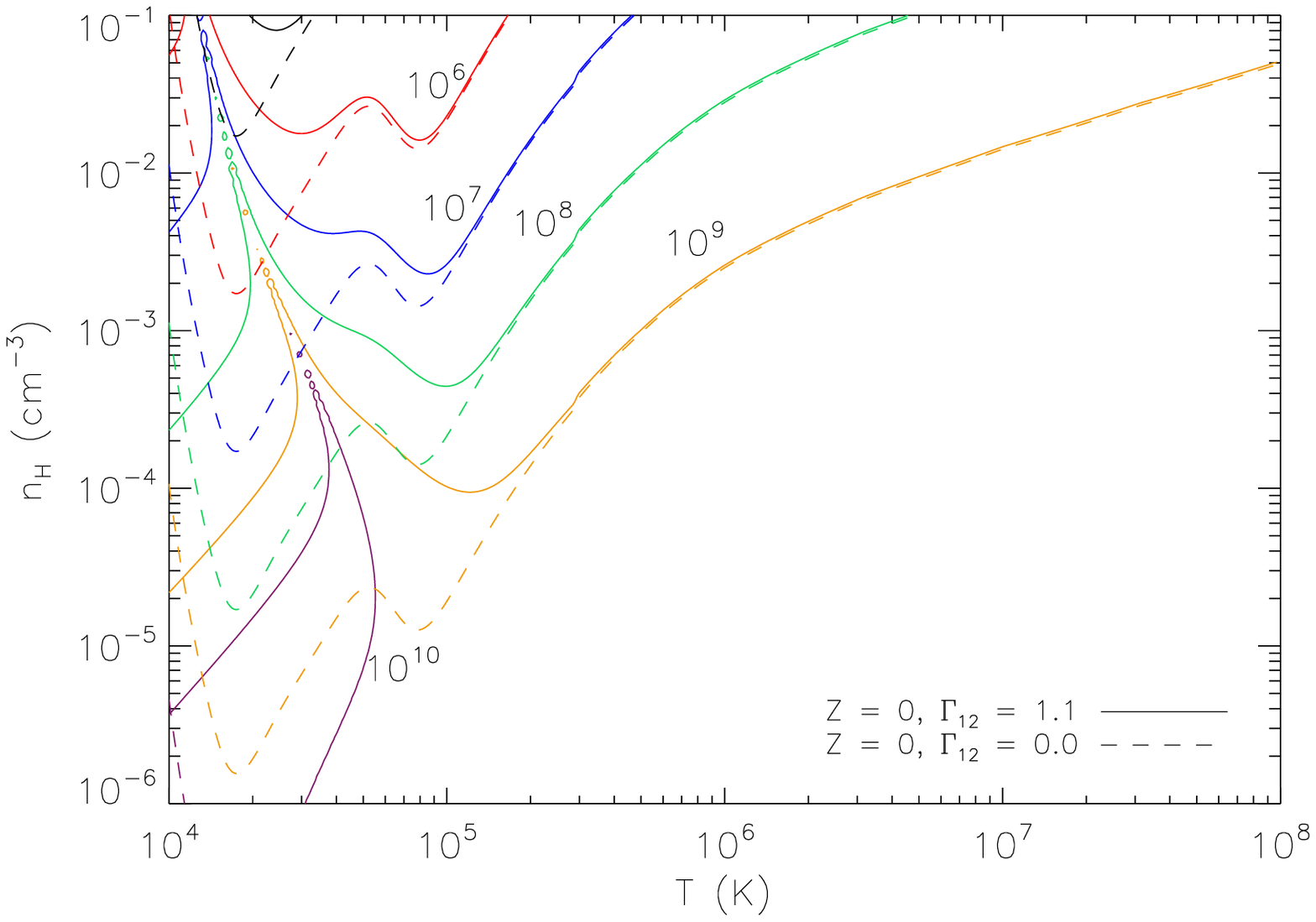}
\includegraphics[width=84mm]{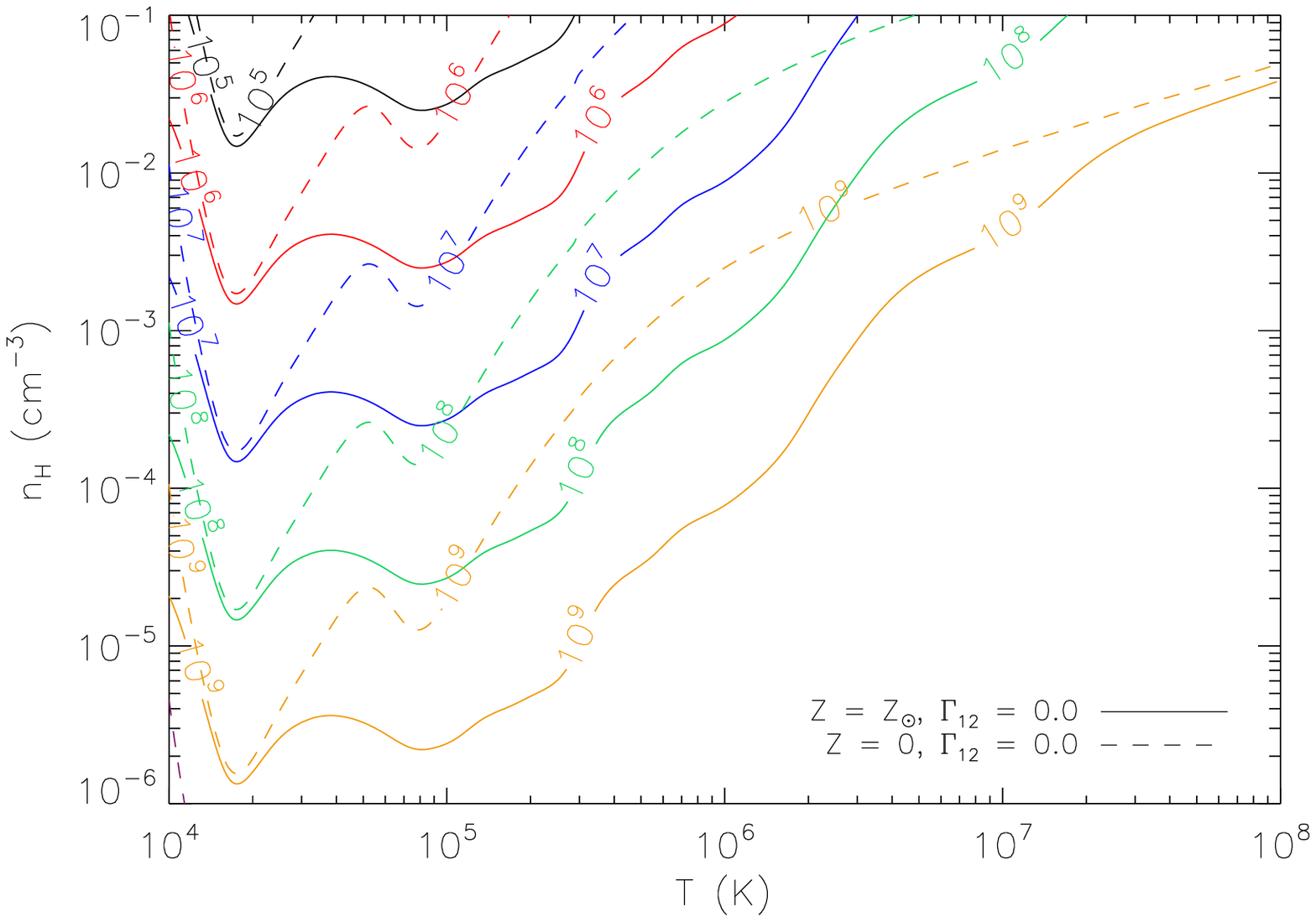}
\includegraphics[width=84mm]{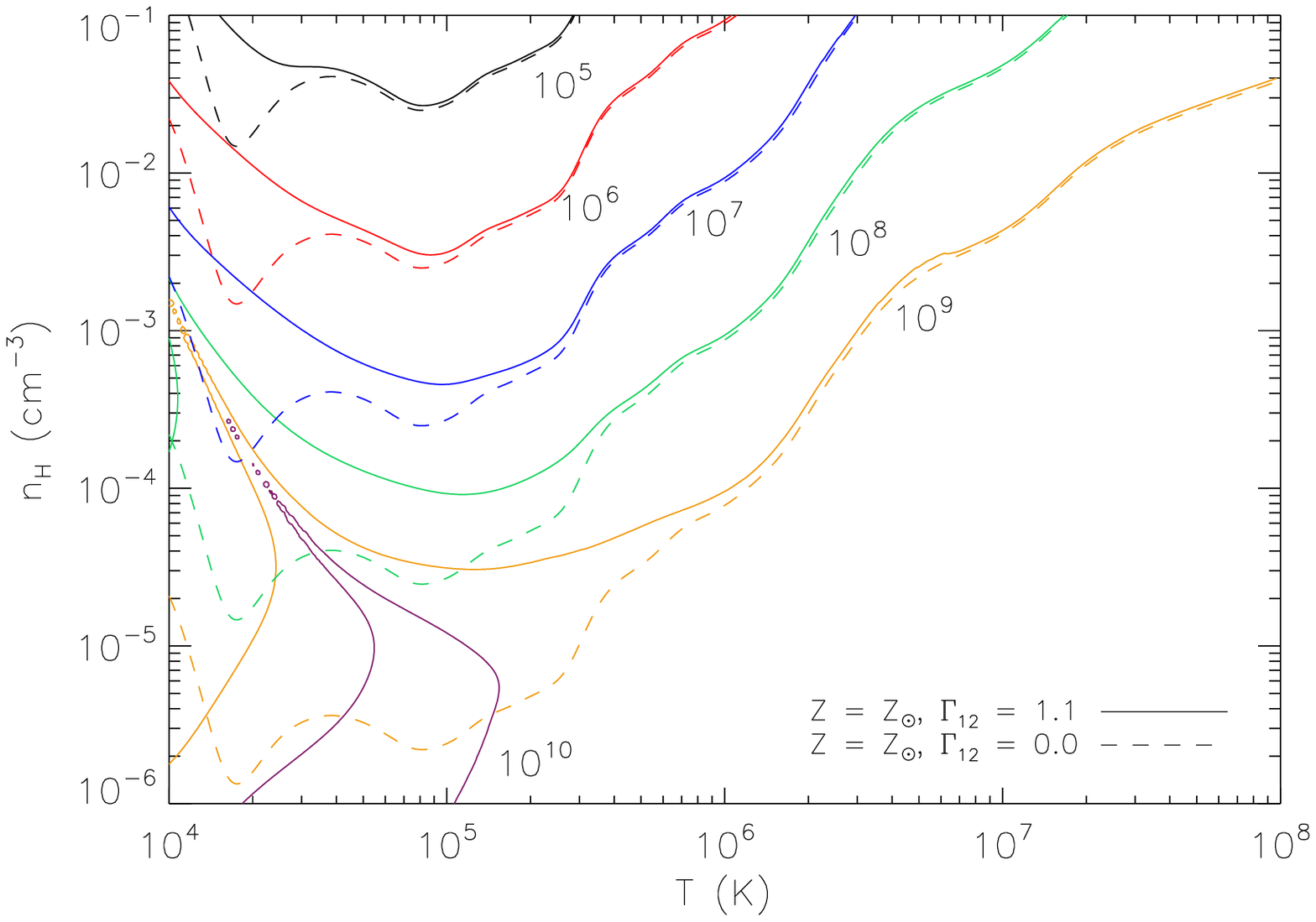}
\includegraphics[width=84mm]{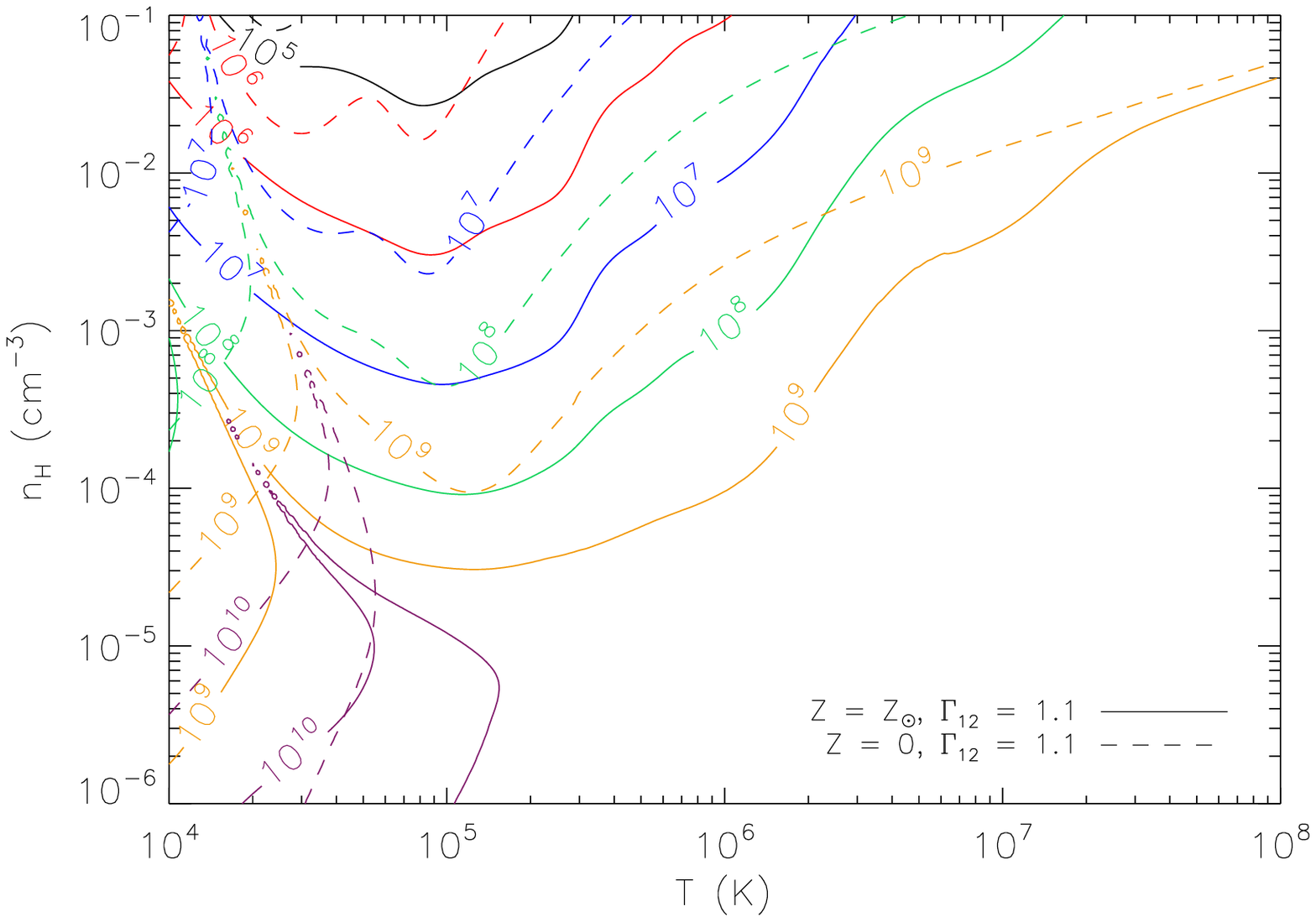}
\caption{Contour plots of the net cooling time in years as a function of
  temperature and total hydrogen number density (note that the mean
  baryon density corresponds to $n_{\rm H} \sim 10^{-5}\,\cm^{-3}$ at
  $z=3$). For temperatures below the equilibrium temperature, the contours indicate 
  negative net cooling times, that is, heating times (see text). 
  The gas in all models is exposed to the $z=3$ CMB. In
  addition, models for which the legend indicates $\Gamma_{12} = 1.1$
  use the $z=3$ HM01 UV/X-ray background. The figure confirms that
  both metals and photo-ionization affect the cooling rates
  significantly, and shows their combined effect.  \emph{Top-left:}
  The effect of photo-ionization for zero metallicity;
  \emph{Top-right:} The effect of metallicity for a purely
  collisionally ionized plasma; \emph{Bottom-left:} The effect of
  photo-ionization for solar metallicity; \emph{Bottom-right:} The
  effect of metallicity for a photo-ionized plasma.
\label{fig:coolingtimecontours}}
\end{figure*}

Figure~\ref{fig:coolingtimecontours} shows how the presence of heavy
elements and radiation affects the radiative cooling time. Each panel
shows contours of constant cooling time in the density-temperature
plane for two models. The top panels illustrate well known results, while
the bottom panels demonstrate the importance of radiation on the
cooling due to metal lines.

Here and throughout the cooling time refers to the absolute value of
the net radiative cooling time at a fixed hydrogen density,
\begin{equation}
t_{\rm cool} \equiv {T \over dT/dt} = {{3 \over 2} n k T \over \left |
  \Lambda_{\rm heat} - \Lambda_{\rm cool}\right |}.
\end{equation}
For temperatures below the thermal equilibrium temperature, $t_{{\rm cool}}$
corresponds to a heating timescale. The calculation includes 
the CMB and, optionally, the HM01 background
at redshift three. 

Note that adiabatic cooling due to the expansion of the Universe is
not included, but would dominate over radiative cooling for
sufficiently low densities. For example, the universal Hubble
expansion (appropriate for densities around the cosmic mean,
i.e.,\ $n_{\rm H} \sim 10^{-5}\,\cm^{-3}$ at $z=3$) corresponds to an
adiabatic cooling time $t_{\rm cool, adiab} = 1/(2H) \approx 1.6\times
10^9$~yr at $z=3$.

The top-left panel shows that for a metal-free\footnote{We use $n_{\rm
    He}/n_{\rm H} = 0.083$ for primordial abundances.} gas, ionizing
radiation drastically reduces the cooling rates for $10^4\,\K < T \la
10^5 \K$ \citep{Efstathiou1992}. This happens because the peaks in the
cooling rate due to collisional excitation of neutral hydrogen and
singly ionized helium (followed by Ly$\alpha$ emission) are removed
when the gas is photo-ionized. 

In the presence of ionizing radiation (solid curves)
there are actually two contours for each net cooling time, corresponding
to net heating (left of the thermal equilibrium asymptote) and net cooling
(right of the equilibrium asymptote). These two contours nearly 
merge\footnote{The fact that the contours really do merge and then
  disappear at high densities is due to the limitations of our
  plotting package and the finite resolution of our grid.} near the
(density-dependent) thermal equilibrium temperature, slightly above
$10^4\,\K$, where the net cooling time goes
to infinity. Below this temperature, the heating time is dictated by
the photo-heating rate. 

For $T\gg 10^5\,\K$ photo-ionization has no
effect on a metal-free gas because the plasma is already fully ionized
by collisional 
processes. In this regime the plasma cools predominantly via the
emission of Bremsstrahlung and/or inverse Compton scattering of CMB
photons. The latter process is the dominant cooling mechanism for most
of the baryons at high redshift ($z > 7$). In the plot inverse Compton
cooling off the CMB dominates the radiative cooling rate at low
densities and high temperatures, but the corresponding cooling time
exceeds the Hubble time.

For diffuse, intergalactic gas (i.e., for density contrasts $\delta
\ll 10^2$, corresponding to $n_{\rm H} \ll 10^{-3}\,\cm^{-3}$ at $z=3$, much
lower than expected in virialized 
objects, e.g.,\ \citealt{Coles2002}) of primordial composition, radiation
increases the cooling time at 
$10^5\,\K$ by at least an order of magnitude and by much more at lower
temperatures. Since the cooling times in this regime are comparable to
the Hubble time, radiation will have a large effect on the fraction of
the baryons that are hot. At densities corresponding to collapsed
objects ($\delta \ga 10^2$ or $n_{\rm H} \ga 10^{-3}\,\cm^{-3}$ at
$z=3$), the increase in the cooling time is generally 
smaller, although it can still easily be an order of magnitude at
temperatures as high as $10^{4.5}\,\K$. 

The top-right panel shows that heavy elements strongly increase the
cooling rate of a collisionally ionized plasma for $10^4\,\K \ll T \la
10^7\,\K$ \cite[e.g.,][]{Boehringer1989}. Comparing the model with
primordial abundances (dashed contours) to the one assuming solar
metallicity (solid contours), we see that the cooling times typically
differ by about an order of magnitude. The presence of metals allows
radiative cooling through collisional excitation of a large number of
ions at a variety of temperatures. For $T> 10^7\,\K$ the difference is
smaller because there are few lines to excite since most elements are
collisionally ionized to a very high degree. Bremsstrahlung is the
dominant cooling mechanism at these very high temperatures.

The bottom-left panel demonstrates that ionizing radiation also
strongly reduces the cooling rates when heavy elements dominate the
cooling. This happens for the same reason as in the primordial
case. The radiation field ionizes the plasma to a higher degree than
it would be in CIE. Hence, the ions that are typically collisionally
excited are not present, reducing the cooling rates.

Note that a similar thing happens when a collisionally ionized plasma
is cooling more quickly than it can recombine. In that case the
ionization balance shifts out of equilibrium leaving the gas too
highly ionized for its temperature. As for photo-ionization, the
associated reduction of the number of bound electrons with excitation
energies low enough to be collisionally excited, reduces the cooling rate 
\cite[e.g.,][]{Kafatos1973,Shapiro1976,Schmutzeler1993,Sutherland1993}. 

Comparing the models with (solid contours) and without ionizing
radiation (dashed contours), we see that for densities characteristic
of the diffuse IGM ($\delta \ll 10^2$ corresponding to $n_{\rm H} \ll 10^{-3} 
 \,{\rm cm}^{-3}$ at $z=3$), radiation significantly reduces the cooling rates
for $T \la 10^6\,\K$ and that the reduction typically exceeds one
order of magnitude for $T \la 3\times 10^5\,\K$. For densities $n_{\rm
  H} \ga 10^{-3}\,\cm^{-3}$ the cooling rates are only suppressed
substantially for $T < 3\times 10^4\,\K$. However, we stress that what matters
here is the ionization parameter. Hence, a stronger radiation field
will affect the cooling rates up to higher densities. Gas with a
density characteristic of collapsed objects will typically be close to
sources of ionizing radiation and may thus be exposed to a radiation
field that is more intense than the meta-galactic UV background.

Finally, the bottom-right panel shows that metals strongly increase
the cooling rates in the presence of an ionizing radiation field, as
was the case for a collisionally ionized gas (top-right panel). For
solar metallicity the cooling rate at $T < 10^7\,\K$ is typically an
order of magnitude higher than for a primordial composition.

Note that for very low densities ($n_{\rm H} \la 10^{-5}\,\cm^{-3}$)
metals increase the equilibrium temperature (towards the left side of
the region where the net cooling time is above $10^{10}$ years)
because their presence boosts the photo-heating (via oxygen and
iron\footnote{The increased He$/$H ratio also boosts the photo-heating
  rate, but its contribution is smaller than that of the metals.})
but does not significantly affect the cooling because it is dominated 
by Compton cooling off the CMB.

\section{Effect on the WHIM}
\label{sec:whim}
The fraction of gas that has been shock-heated to temperatures of
$10^5\,\K ~\la ~T < 10^7\,\K$ is currently of great interest, mainly
because this so-called warm-hot intergalactic medium (WHIM) is hard to
detect, yet may contain a large fraction of the baryons in the
low-redshift Universe (e.g., \citealt{Cen1999}). There are two reasons
why the WHIM becomes more important at lower redshift. First, as
structure formation progresses, larger structures form, leading to
stronger gravitational accretion shocks and a greater fraction of the
baryons are heated to temperatures in the WHIM range. Second, as the
universe expands, the density of the diffuse gas decreases
as\footnote{In reality the density will decrease slightly less fast
  with time if the gas is overdense and collapsing.} $(1+z)^3$ and the
cooling time due to collisional processes (which dominate for $z < 7$)
will thus increase as $(1+z)^{-3}$. Hence, the cooling time increases
faster than the Hubble time and more and more of the shock-heated gas
is unable to cool.

\begin{figure}
\includegraphics[width=84mm]{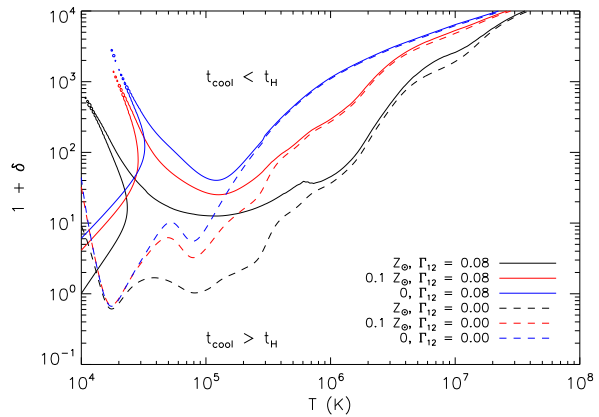}
\caption{Contours show where the net cooling time equals the Hubble time
  in the density contrast - temperature plane at $z=0$. From top to
  bottom, the different colours correspond to primordial composition
  (blue), a metallicity of ten percent solar (red), and solar
  abundances (black). Dashed contours are for collisional ionization
  only, while solid contours are for gas exposed to the HM01 model for
  the $z=0$ UV/X-ray background. Each solid contour comprises two components,
  corresponding to net heating (low temperature component) and net
  cooling (high temperature component), respectively, which
  merge near the (density-dependent) thermal equilibrium temperature.
  Both photo-ionization and metallicity
  determine whether gas is able to cool and it is therefore crucial to
  take both into account when predicting the fraction of the baryons
  that reside in the WHIM.
\label{fig:whimcooling}}
\end{figure}

Since the cooling times are of order the Hubble time for much of the
WHIM, the precise values of the cooling rate is particularly
important. Because the cooling of the WHIM tends to be dominated by
line radiation, because its density is low \cite[$\delta \sim 10-10^2$; e.g.,][]{Bertone2008},
and because the WHIM gas may well be enriched to values of 10\% of
solar or higher, both metals and photo-ionization by the UV background
may be important. Figure~\ref{fig:whimcooling} demonstrates that this
is indeed the case.

Figure~\ref{fig:whimcooling} shows a contour plot of the cooling time
in the density contrast-temperature plane for redshift $z=0$. Each
contour corresponds to the same net cooling time, namely the Hubble
time. Dashed contours are for a purely collisionally ionized gas,
while the solid contours include photo-ionization by the $z=0$
UV/X-ray background radiation from galaxies and quasars, which
corresponds to a hydrogen photo-ionization rate $\Gamma_{12} = 0.08$
and an ionization parameter $U = 9 \times 10^{-7}/n_{\rm
  H}$. Metallicity increases from the top down from zero (i.e., 
primordial abundances) (blue contours), to 10\% solar (red) to solar
(black), as indicated in the figure. Gas above a given contour is able
to cool, while gas below it will remain hot.

For a primordial composition (blue contours), turning on the ionizing
radiation raises the density contrast above which the gas can cool
from $\delta \la 10$ to $\delta \sim 10^2$ for $T < 10^5\,\K$. For
solar metallicity, the UV radiation becomes important for $T<10^6\,\K$
and raises the critical density required for cooling within a Hubble
time by about an order of magnitude for $T \la 10^5\,\K$. Increasing
the metallicity from zero to solar decreases the critical density
contrast by about an order of magnitude over the full range of WHIM
temperatures. Clearly, both photo-ionization and metals are important
for the thermal evolution of the WHIM.

\begin{figure}
\includegraphics[width=84mm]{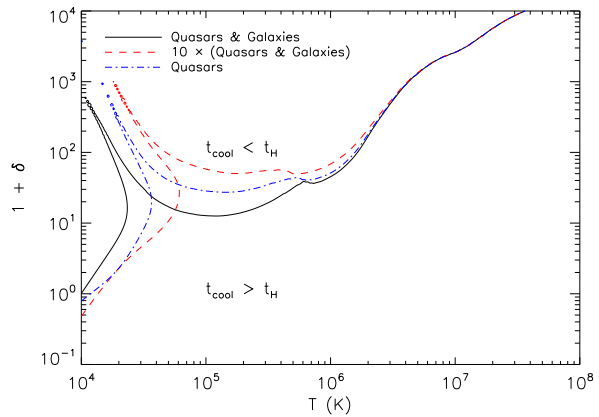}
\caption{Contours show where the net cooling time equals the Hubble time
  in the density contrast - temperature plane for gas of solar
  metallicity at $z = 0$. The solid contour is for the HM01 model for
  the $z=0$ UV/X-ray background from galaxies and quasars (model QG),
  the dashed contour is for a radiation field with the same spectral
  shape as model QG, but 10 times the intensity, and the dot-dashed
  contour is for the HM01 quasars only radiation field, scaled to the
  same hydrogen ionization rate as model QG. Each contour comprises
  two components, 
  corresponding to net heating (low temperature component) and net
  cooling (high temperature
  component), respectively, which 
  merge near the (density-dependent) thermal equilibrium
  temperature. 
\label{fig:uvbqso}}
\end{figure}

Figure~\ref{fig:uvbqso} shows how the intensity and the spectral shape
of the ionizing radiation affect the results for solar abundances. The
solid contour indicates the density contrasts and temperatures for
which the cooling time equals the Hubble time for the $z=0$ HM01 model
for the UV/X-ray radiation from quasars and galaxies (model QG). It is
identical to the black, solid contour in
Figure~\ref{fig:whimcooling}. The dashed contour shows the $t_{\rm
  cool} = t_H$ contour after we have multiplied the entire HM01
radiation field by a factor of 10 (model 10$\ast$QG). Increasing the
intensity reduces the cooling rates, shifting the contour to higher
densities.

The dot-dashed contour corresponds to the quasar only HM01 model
(model Q), which we have rescaled such as to give the same hydrogen
ionization rate as model QG. Since the average energy of H-ionizing
photons is higher for model Q than for QG, the difference between the
dot-dashed and solid contours reflects the effect of changing the
spectral hardness. Comparing these two contours, we see that, at a
fixed H ionization rate, harder spectra tend to inhibit the cooling
more than softer spectra for temperatures above the equilibrium value
(the $t_{\rm cool}  =t_H$ contour moves to higher densities for harder spectra).
This happens because collisional excitation 
of heavy ions involves electrons with higher ionization
energies than hydrogen. The extra high energy photons will remove many
of those electrons, thus reducing the cooling rates further. For
temperatures below the equilibrium value, however, the net cooling
time actually decreases (the lower-left of the left contour moves
to lower densities for harder spectra) because the photo-heating rates are
larger for harder spectra. 

In summary, for densities and temperatures characteristic of the WHIM
the nature of the ionizing 
radiation field as well as the metallicity of the gas may have a significant
impact. Accurate cooling rates therefore require a correct treatment
of the composition of the gas, the spectral 
hardness, and the radiation intensity. Simple models may result in a
poor estimate of the amount of baryons in the WHIM phase.

\begin{figure*}
\includegraphics[width=84mm]{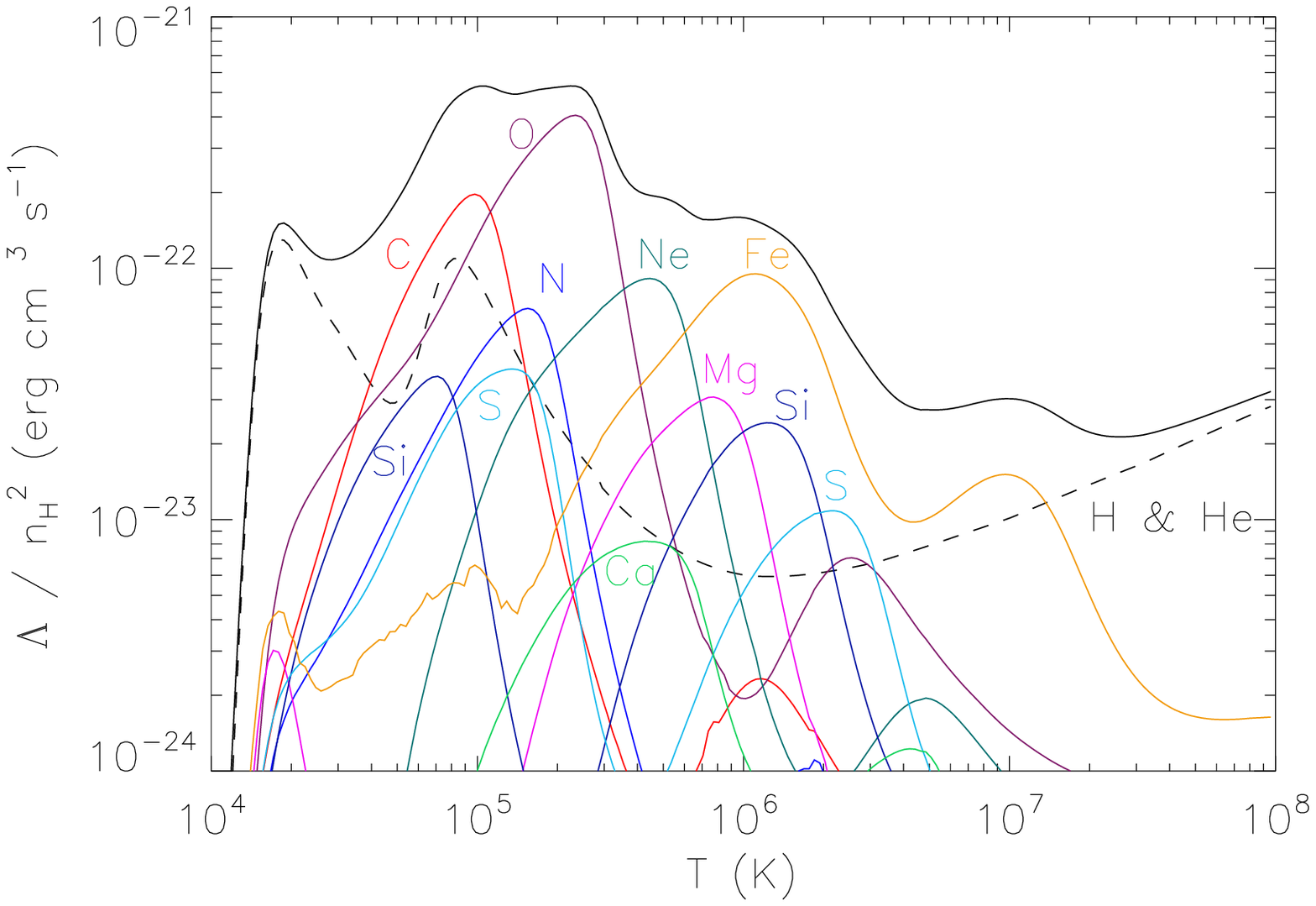}
\includegraphics[width=84mm]{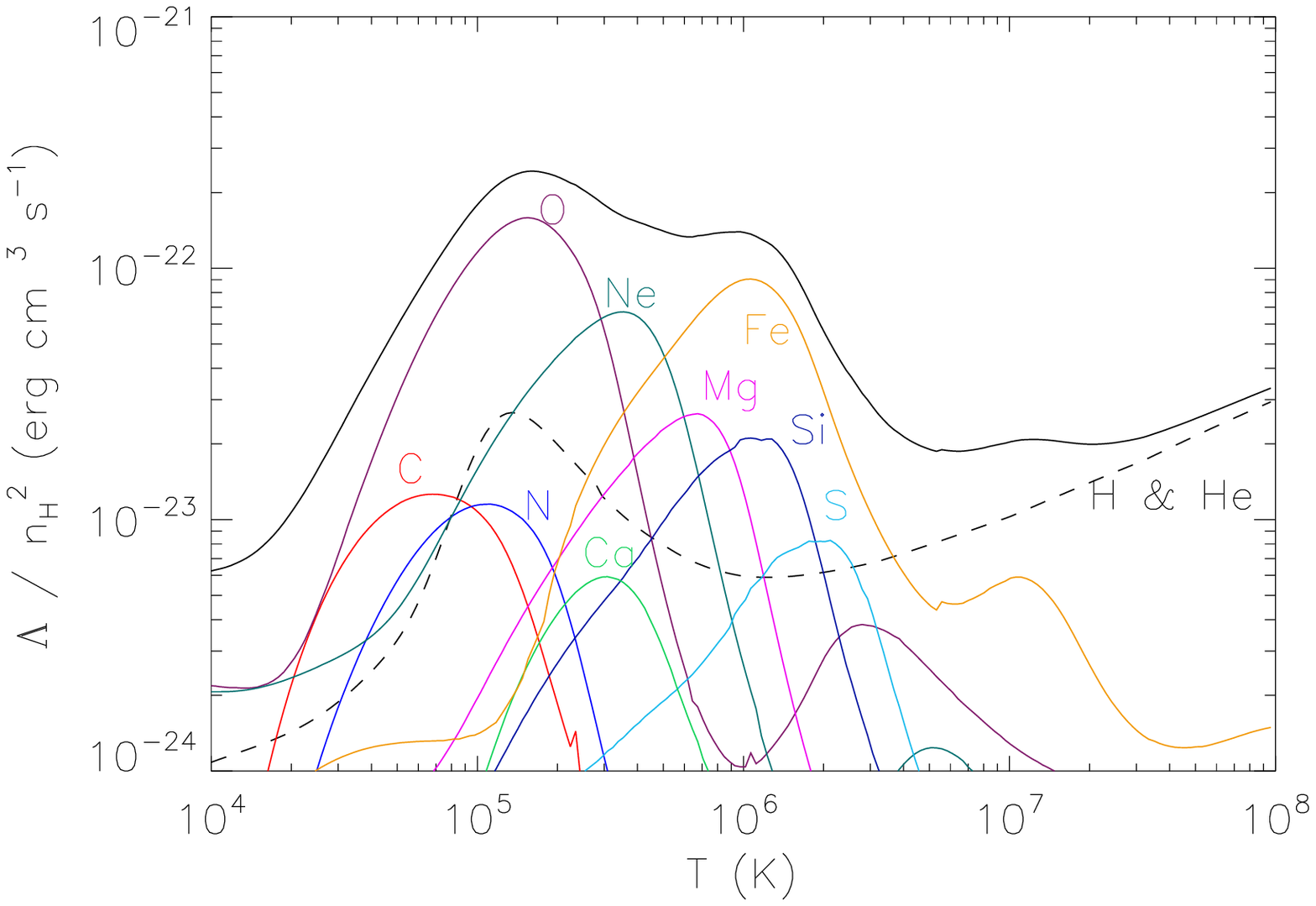}
\caption{Normalized cooling rates as a function of temperature for
  solar abundances, assuming either CIE (\emph{left-hand panel}) or
  photo-ionization equilibrium for $n_{\rm H} = 10^{-4}\,\cm^{-3}$ and
  an optically thin gas exposed to the $z=3$ HM01 model for the
  meta-galactic UV/X-ray background from quasars and galaxies
  (\emph{right-hand panel}). Note that normalized cooling rates are
  independent of the density for CIE, but not for photo-ionization
  equilibrium. The black, solid curve indicates the total
  cooling rate and the thin, coloured, solid curves show the
  contributions from individual elements. The black, dashed curve
  shows the contribution from H and He.
\label{fig:elements}}
\end{figure*}

\section{The relative importance of different elements}
\label{sec:relabund}
For solar abundances, a large number of elements contribute to the
radiative cooling rate. The black, solid curve in the left-hand
panel of Figure~\ref{fig:elements} shows the normalized cooling rate ($\Lambda/n_{\rm H}^2$)
as a function of temperature for a plasma in pure CIE and for solar
abundances, reproducing previous work done on this subject \cite[e.g.,][]{Cox1969,Raymond1976,Shull1982,Gaetz1983,Boehringer1989,Sutherland1993,Landi1999,Benjamin2001,Gnat2007,Smith2008}. The thin, coloured, solid curves show the contributions
from individual elements, while the black, dashed curve shows the
total cooling rate due to H and He. For $T\gg 10^7\,\K$ Bremsstrahlung
(i.e.,\ H \& He) dominates, but at lower temperatures line radiation is
most important. Going down from $10^7\,\K$ to $10^4\,\K$ the cooling
rate is successively dominated by iron, neon, oxygen, carbon, and
hydrogen.

If the cooling rate is dominated by a single element, as is for
example the case for oxygen at $T\approx 2\times 10^5\,\K$ and for
iron at $T\sim 10^6\,\K$, then the total cooling rate will be
sensitive to the relative abundances of those elements. For instance,
since $[{\rm O}/{\rm Fe}]$ is observed to vary with environment by
factors of two or so at a fixed metallicity
\cite[e.g.,][]{Shetrone2003}, we can expect similar variations in the
cooling rates.

We next turn on the HM01 ionizing background and illustrate the
results for $n_{\rm H}=10^{-4}\,\cm^{-3}$ and $z = 3$ in the
right-hand panel of Figure~\ref{fig:elements} (recall that the
left-hand panel is independent of both density and redshift). For this
figure we have 
excluded Compton cooling off the CMB to isolate the impact of the
ionizing radiation and because the former is only important at high
redshift. We can see that photo-ionization affects some elements more
than others. As we have seen before, the effect is stronger for lower
temperatures. Although we show the results for only a single density
here, we note that the importance of photo-ionization increases with
decreasing density.

Comparing the cooling rates including photo-ionization (right-hand
panel of figure~\ref{fig:elements}) to those for CIE (left-hand panel of
figure~\ref{fig:elements}) shows that photo-ionization increases the
relative importance of oxygen and decreases that of carbon, helium,
and especially hydrogen. It is also clear that many of the peaks of the
various elements shift to lower temperatures when an ionizing
radiation field is present. This shift occurs because a photo-ionized
gas is overionized for its temperature compared to a collisionally
ionized plasma. If the ion fractions peak at lower temperatures, then
so will the cooling rates due to collisional excitation of those ions.

This last figure illustrates the central result of this work:
photo-ionization changes both the total cooling rates and the relative
importance of individual elements. For a more complete visualization
of this point, we kindly refer the interested reader to our web
site\footnote{\texttt{http://www.strw.leidenuniv.nl/WSS08/}.}, where we
host a number of videos, plots, and the tables themselves for
download.

\section{Discussion}
\label{sec:discussion}
Radiative cooling is an essential ingredient of hydrodynamical models
of a wide range of astrophysical objects, ranging from the IGM to
(proto-)galaxies and molecular clouds.  While numerical simulations of
objects with a primordial composition often compute non-equilibrium
radiative cooling rates explicitly and sometimes even include the
effect of ionizing background radiation, the treatment of cooling of
chemically enriched material is typically much more approximate. For
example, simulations of galaxy formation typically either ignore
metal-line cooling altogether or include it assuming pure CIE. In
addition, the abundances of all heavy elements are typically scaled by
the same factor (the metallicity) (but see \citealt{Martinez2008} and
\citealt{Maio2007} for recent exceptions). In this simplified
treatment metal-line cooling depends only on temperature and
metallicity, allowing straightforward interpolation from pre-computed
two-dimensional tables.

We have used \textsc{cloudy} to investigate the effects of heavy
elements and ionizing radiation on the radiative cooling of gas with
properties characteristic of (proto-)galaxies and the IGM, i.e.,
optically thin gas with densities $n_{\rm H} \la 1\,\cm^{-3}$ and
temperatures $T\ga 10^4\,\K$, assuming ionization equilibrium. We
presented a method to incorporate radiative cooling on an
element-by-element basis including photo-ionization by an evolving
UV/X-ray background, using precomputed tables, which for heavy
elements are functions of density, temperature, and redshift and for
H\&He (which must be considered together because they are important
contributors to the free electron density) depend additionally on the
He/H ratio. Using the 11 elements H, He, C, N, O, Ne, Mg, Si, S, Ca,
and Fe, the redshift $z=0$ median absolute errors in the net cooling
rate range from 0.33\%, at $Z = 0.1 Z_{\odot}$ to 6.1\% for the
extreme metallicity $Z= 10 Z_{\odot}$, and the errors are smaller for
higher redshifts. 

The tables as well as some scripts that illustrate how to use them are
available from the following web site:
\texttt{http://www.strw.leidenuniv.nl/WSS08/}. We also include tables
for solar relative abundances which can be used if metallicity, but
not the abundances of individual elements are known, as in equation
(\ref{eq:Zmethod}). This web site also contains a number of videos
that may be helpful to gain intuition on the importance of various
parameters on the cooling rates.

We confirmed that, assuming CIE, heavy elements greatly enhance the
cooling rates for metallicities $Z 
\ga 10^{-1}~Z_\odot$ and temperatures $T \la 10^7\,\K$. We
demonstrated that this remains true in the
presence of photo-ionization by the meta-galactic UV/X-ray
background. 

The background radiation removes electrons that would otherwise be
collisionally excited, thus reducing the cooling rates. The effect is
stronger for higher ionization parameters (i.e.,\ higher radiation
intensities or lower densities) and if the spectral shape of the
radiation field is harder. Considering only the meta-galactic
radiation field, which provides a lower limit to the intensity of the
radiation to which optically thin gas may be exposed, the reduction of
the metal-line cooling rates becomes important below $10^6\,\K$ for
ionization parameters $U \ga 10^{-1}$ and below $10^5\,\K$ for $U \ga
10^{-3}$ (note that for the HM01 background $U=9\times 10^{-7}/n_{\rm
  H}$ and $2\times 10^{-5}/n_{\rm H}$ at $z=0$ and $z=3$, respectively).  

As an example of the potential importance of including the effects of
both photo-ionization and heavy elements, we considered the so-called
warm-hot intergalactic medium (WHIM), which is thought to contain a
large fraction of the baryons at 
redshifts $z < 1$. We demonstrated that the overdensities for which
gas at typical WHIM metallicities ($Z\sim 10^{-1}~Z_\odot$) and
temperatures ($T\sim 10^5 - 10^7\,\K$) can cool within a Hubble time,
can shift by an order of magnitude depending on whether
photo-ionization and metal-line cooling are taken into account. Hence,
photo-ionization of heavy elements may have important consequences for
predictions of the amount of matter contained in this elusive gas
phase.  

Because chemical enrichment happens in a number of stages, involving a
number of processes with different timescales, the relative abundances
of the heavy elements varies with redshift and environment by factors
of a few. Hence, computing cooling rates on an
element-by-element basis rather than scaling all elements by the
metallicity, will change the cooling rates by factors of a few. The
difference is therefore typically somewhat smaller than the effect of
neglecting metals or photo-ionization altogether, but still highly
significant.

While it was known that different elements dominate the cooling for
different temperatures in CIE, we showed that photo-ionization both
shifts the peaks due to individual elements to smaller temperatures
and reduces their amplitude. Note that since photo-ionization
over-ionizes the gas, this effect is similar (but not equivalent to)
that found in non-equilibrium calculations without ionizing radiation
\cite[e.g.,][]{Sutherland1993, Gnat2007}. Because the importance of
photo-ionization depends on the ionization parameter, the relative
contributions of individual elements exposed to a fixed ionizing
radiation field depends also on the gas density.

Would dropping our assumption of ionization equilibrium have a large
effect on the cooling rates?  Ionizing radiation results in a plasma
that is overionized relative to its temperature. Its effect is
therefore similar to that of non-equilibrium ionization following
rapid cooling (i.e., if the cooling time is shorter than the
recombination times of the ions dominating the cooling, see
e.g.,\ \citealt{Kafatos1973,Shapiro1976}). We therefore anticipate that
the effect of non-equilibrium ionization will be much smaller for our
cooling rates than for those that assume CIE.

The assumptions that the gas is optically thin and exposed only to the
meta-galactic background radiation are likely to be more important
than the assumption of ionization equilibrium, particularly since
non-equilibrium collisional cooling rates only differ from those
assuming CIE by factors of a few or less
\cite[e.g.,][]{Schmutzeler1993,Sutherland1993,Gnat2007}. For column
densities $N_{\rm HI} > 10^{17}~\cm^{-2}$ self-shielding becomes
important and only part of the H-ionizing radiation will penetrate the
gas cloud, which would particularly affect the cooling rates for $T\la
10^5\,\K$. At higher temperatures line cooling is dominated by heavier
ions, which can only be ionized by higher energy photons and which
therefore remain optically thin up to much higher column densities.
This is because the photo-ionization cross sections of H and He drop
rapidly with increasing frequency for energies exceeding their
ionization potentials. Moreover, for $T\gg 10^4~\K$ hydrogen is
collisionally ionized to a high degree and consequently the optical
depth for ionizing radiation will be significantly reduced.

It is, however, far from clear that high column densities would reduce
the effect of radiation. For self-shielded clouds the cooling
radiation may itself be trapped, providing a source of ionizing
radiation even in the absence of an external one
\cite[e.g.,][]{Shapiro1976,Gnat2007}. Moreover, gas clouds with columns
that exceed $10^{17}~\cm^{-2}$ are on average expected to be
sufficiently close to a galaxy that local sources of ionizing
radiation dominate over the background \citep{Schaye2006,Miralda2005}.

Ultimately, these issues can only be resolved if non-equilibrium
cooling rates are computed including radiative transfer and if
the locations of all relevant sources of ionizing radiation are known.  It
will be some time before it is feasible to carry out such a calculation
in, say, a cosmological hydrodynamical simulation. In the mean time,
we believe that our element-by-element calculation of the equilibrium
cooling rates for an optically thin gas exposed to the CMB and an
evolving UV/X-ray background provides a marked improvement over
earlier treatments. In future publications we will present
cosmological, hydrodynamical simulations using these cooling rates.

\section*{Acknowledgments}
We are grateful to the anonymous referee whose helpful comments
greatly improved the manuscript. We are also grateful to Gary Ferland
for help with \textsc{cloudy} and to Brent Groves for help with
\textsc{mappings iii}. We would also like to thank Tom Abel and the
members of the OWLS collaboration for discussions. In particular, we
are grateful to Tom Theuns for showing us the benefits of HDF5. This
work was supported by Marie Curie Excellence Grant
MEXT-CT-2004-014112.

\bibliographystyle{mn2e} \bibliography{ms}
\label{lastpage}

\end{document}